\documentclass[aps,prl,twocolumn,groupedaddress,amsmath,amsfonts,amssymb]{revtex4}

\usepackage{graphicx}
\usepackage{bm}

\newcommand{\bra}[1]{\langle #1|}
\newcommand{\ket}[1]{|#1\rangle}

\newcommand{\mm}[1]{\mathrm{#1}}

\begin{document}

\title{Nuclear State Preparation via Landau-Zener-St\"uckelberg transitions in Double Quantum Dots}

\author{Hugo Ribeiro}
%\email{hugo.ribeiro@uni-konstanz.de}

\author{Guido Burkard}
%\email{guido.burkard@uni-konstanz.de}

\affiliation{Department of Physics, University of Konstanz, D-78457 Konstanz, Germany}


%\date{\today}

\begin{abstract}

We theoretically model a nuclear-state preparation scheme that increases the coherence
time of a two-spin qubit in a double quantum dot. The two-electron system is tuned
repeatedly across a singlet-triplet level-anticrossing with alternating slow and rapid
sweeps of an external bias voltage.  
Using a Landau-Zener-St\"uckelberg model, we find that in addition to a small nuclear
polarization that weakly affects the electron spin coherence, the slow sweeps are only
partially adiabatic and lead to a weak nuclear spin measurement and a nuclear-state
narrowing which prolongs the electron spin coherence. This resolves some open problems
brought up by a recent experiment \cite{reilly_science}.
Based on our description of the weak measurement, we simulate a system with up to $n=200$
nuclear spins per dot. Scaling in $n$ indicates a stronger effect for larger $n$.

\end{abstract}

\maketitle

\textit{Introduction---}
Since the electron spin in quantum dots (QDs) has been proposed as a qubit
\cite{loss_divicenzo_proposaL_QD}, much progress has been made to develop reliable
semiconductor devices, mostly with GaAs, in which only a single electron can be confined
and its spin can be controlled \cite{hanson_review}.
Despite the achievement of reliable control, spin decoherence due to the hyperfine
interaction with the surrounding nuclear spins remains a major problem
\cite{khaetskii_e-spin_decoherence}. 
In double QDs, where the singlet ($\mm{S}$) and one of the triplet ($\mm{T}_0$) states
can be chosen as an effective two-level system to implement a qubit
\cite{levy_QC_2e,taylor_nature_phys,hanson_burkard}, it has been shown that the
probability, averaged over many runs of the experiment, to find the system in the singlet
state at time $t>0$, having prepared a singlet at time $t=0$, is a decaying oscillating
function $\propto\cos(Jt)\exp(-(t/T_2^*)^2)$ \cite{petta_science,coish_s-t_decoherence},
with singlet-triplet exchange splitting $J$ and decoherence time $T_2^*$. 

An important goal in the quest to overcome decoherence of spin qubits in solid-state
devices is to find mechanisms that allow for an increase in $T_2^*$. Spin echo has been
used to reveal $T_2 \approx \mathrm{\mu s} \gg T^{*}_2 $ \cite{petta_science} which sets
the scale that can be achieved by nuclear-state preparation. In principle, it is possible
to reduce the nuclear fluctuations, thus prolonging $T_2^*$, by projecting the nuclear
state into (approximate) Overhauser eigenstates with either electrical or optical means
\cite{klauser_nucl_spin_narr, stepanenko_enh_spin_coh_opt} or by polarizing the nuclear
spins \cite{burkard_polarization, petta_polarization}. However, a sizable enhancement of electron spin
coherence would only be realized for a polarization of more than 99\% \cite{coish_99},
which so far has not been achieved. 

\begin{figure} \includegraphics[width=0.48\textwidth]{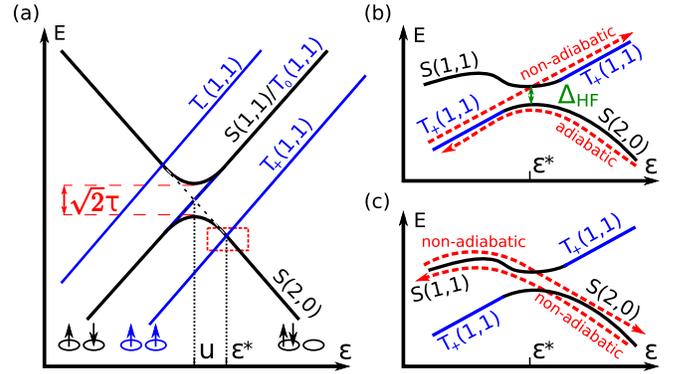}
\caption{(color online) (a) Energy diagram for the relevant states of the double QD as a
function of the bias $\varepsilon$. 
(b) The hyperfine interaction allows for flip-flop processes that open a gap $\Delta_{\rm
HF}$ at the $\mm{S}$-$\mm{T}_{+}$ crossing. 
Driving the system slowly from $\mm{S}(2,0)$ can result either in an adiabatic (b) or a
non-adiabatic (c) transition.  Driving the system back very fast results in a
non-adiabatic passage to two distinguishable charge states.} \label{energy_diagram}
\end{figure}

Reilly \textit{et al.} \cite{reilly_science} have experimentally studied adiabatic
transitions of two electrons in a pair of tunnel-coupled QDs from a spin singlet
$\mm{S}(2,0)$, with both electrons in the same QD and total spin $0$, across an energy level
anti-crossing to a spin triplet $\mathrm{T}_{+}(1,1)$, with one electron in each QD and
total spin $1$ (see Fig.~\ref{energy_diagram}). Due to angular momentum conservation, the
hyperfine-induced transition $\mathrm{S}$-$\mathrm{T}_{+}$ involves an electron spin flip
accompanied by a nuclear spin flop.
In the experiment, the process was repeated many times with intermediate fast resetting
to the singlet state ($200\,\mm{ns}$). If the slow $\mm{S}$-$\mm{T}_+$ transition was
fully adiabatic and the nuclear spin polarization sufficiently long-lived, this cycle
should allow for complete nuclear polarization as the number of cycles becomes comparable
to the number of nuclear spins in the QDs (typically about $10^5$ to $10^6$).
However, a polarization of only about 1\% was achieved. Nevertheless, the coherence time
measured in the $\mm{S}$-$\mm{T}_0$ subspace was improved by a factor of up to 70
\cite{reilly_science} which can be attributed to the preparation of a suitable nuclear
state.
A previous theoretical model has been used to calculate the evolution of up to $36$
nuclear spins per dot for an initial mixed state with \emph{fixed} angular momentum per dot
\cite{ramon_hu}.

Here, we propose a theoretical model of nuclear state preparation taking into account all
possible angular momenta, up to $200$ nuclear spins per QD, and the possibility of
\textit{partially adiabatic} transitions. Such transitions can explain the smallness of
the nuclear polarization generated while allowing for a weak measurement
\cite{aharonov_wm} of the nuclear state. The repeated weak measurements at each cycle
lead to nuclear state narrowing and, as a consequence, to a prolongation of the spin
singlet-triplet coherence. The main feature of our theoretical model is the description
of the partially adiabatic $\mathrm{S}$-$\mathrm{T}_{+}$ transition by the
Landau-Zener-St\"uckelberg (LZS) theory \cite{landau32,zener32,stuckelberg32}. 
Due to the high degeneracy of the nuclear spin state and the finite sweep time, we use
suitable generalizations of the LZS model \cite{vasilev_deg_lz, vitanov_time_lz}. 

%\section{Model}
\textit{Model---}
We derive the Hamiltonian describing the partially adiabatic dynamics at the
$\mathrm{S}$-$\mathrm{T}_{+}$ anti-crossing. We start with the spin-preserving part $H_0$
of the Hamiltonian, with $[H_0,S_z]=0$, describing the coupling between two electrons in
a double QD in a magnetic field $B$,
\begin{equation}
\begin{aligned}
H_0 =& \sum_{i \sigma} \left(\varepsilon_i +
\frac{1}{2}g^{*}\mu_{\mathrm{B}}B\sigma\right) c_{i\sigma}^{\dag} c_{i\sigma} \\
&+ u\sum_{i} c_{i\uparrow}^{\dag}c_{i\uparrow}c_{i\downarrow}^{\dag}c_{i\downarrow}
+ \tau\sum_{\sigma}\left(c_{1\sigma}^{\dag} c_{2\sigma} + \mathrm{h.c}\right),
\end{aligned}
\label{H_trap}
\end{equation}
where $g^{*}$ denotes the effective Land\'e g-factor ($-0.44$ for GaAs),
$\mu_{\mathrm{B}}$ the Bohr magneton, and the indices $i=1,2$ and
$\sigma=\uparrow,\downarrow$ label the dot number and spin. The first term is the
single-particle energy of the confined electrons, the second accounts for 
the Coulomb energy $u$ of two electrons on the same QD, and the last for 
the electron tunneling with strength $\tau$ between the dots.

The diagonalization of the first two terms of Eq.~\eqref{H_trap} leads to the relevant
states of a double QD: the singlets $\mathrm{S}(0,2)$, $\mathrm{S}(2,0)$,
$\mathrm{S}(1,1)$ and triplets $\mathrm{T}_{\pm,0}(1,1)$, where $(l,r)$ indicates the
number of electrons in the (\emph{left}, \emph{right}) dot. The other states can be
neglected as they have energies much higher than those considered here. The degeneracy of
the singlets $\mathrm{S}(2,0)$ and $\mathrm{S}(1,1)$ at $\varepsilon = \pm u$ is lifted
by the inter-dot tunneling, resulting in a splitting of $\sqrt{2} \tau$. The energy
levels as a function of the bias $\varepsilon = \varepsilon_1 - \varepsilon_2$ are shown
in Fig.~\ref{energy_diagram}(a). 
At the degeneracy $\varepsilon = \varepsilon^{*}$ of the singlet $\mm{S}(1,1)$ and the
triplet $\mm{T}_{+}(1,1)$, the hyperfine interaction between the electron spins $
\mathbf{S}_i$ and the nuclear spins $\mathbf{I}_i^k$ opens a splitting $\Delta_{\mm{HF}}$.
The contact hyperfine Hamiltonian is given by $H_{\mathrm{HF}} = \mathbf{S}_1\cdot\mathbf{h}_2 +
\mathbf{S}_2\cdot\mathbf{h}_2$ where $\mathbf{h}_{\mathrm{i}} = \sum_{k=1}^{n(i)} A_i^k
\mathbf{I}_i^k$ is the Overhauser (effective nuclear field) operator. The sum runs over
the $n$ nuclear spins in dot $i$, $A_i^k = v_k \nu_0 \left|\Psi (\mathbf{r}_k)\right|^2$
is the hyperfine coupling constant with the $k$-th nucleus in dot $i$, with $\Psi
(\mathbf{r}_k)$ the electron wave function, $\nu_0$ the volume of the unit cell and $v_k$
the hyperfine coupling strength.  
Introducing $S_i^\pm=S_i^x\pm i S_i^y$ and  $h_i^\pm=h_i^x\pm i h_i^y$, 
we write $H_{\mm{HF}}$ as
\begin{equation}
H_{\mathrm{HF}} = \frac{1}{2}\sum_{i}\left(2 S_i^z h_i^z + S_i^+h_i^- +
S_i^-h_i^+\right).
\label{hyperfine_h}
\end{equation}

Since $\tau \gg \Delta_{\mm{HF}}$, we can use $H = H_0 + H_{\mm{HF}}$ to derive an
effective Hamiltonian for the subspace spanned by $\{\ket{\mathrm{S}, j_1, m_1, j_2,
m_2}\}$ and $\{\ket{\mathrm{T}_{+}, j_1, m_1', j_2, m_2'}\}$, where $j_i$ is the total
nuclear angular momentum in dot $i$ and $m_i$ its projection along $B$,
\begin{equation}
H(\varepsilon)  = \sum_{q\mathbf{\chi}} 
E_{q\mathbf{\chi}}(\varepsilon) 
\ket{q\mathbf{\chi}}\bra{q\mathbf{\chi}}
 + \frac{1}{2}\sum_{i} \left(S_i^+h_i^- + S_i^-h_i^+\right)
\label{Lzhamiltonian}
\end{equation}
where $q= \mm{S}, \mm{T}$, $\ket{\mathbf{\chi}} = \ket{j_1, m_1, j_2, m_2}$, and
$E_{\mathrm{S},\mathbf{\chi}}(\varepsilon) = E_{\mathrm{S}}(\varepsilon)$, and
$E_{\mathrm{T},\mathbf{\chi}}(\varepsilon) = E_{\mathrm{T}}(\varepsilon) +
\langle h^z_1+ h^z_2\rangle_{\mathbf{\chi}}/2 + g^{*}\mu_{\mathrm{B}}B$.  

\textit{Method---}
With a time-dependent bias $\varepsilon = \varepsilon(t)$,
the Hamiltonian \eqref{Lzhamiltonian} is of the form $H = H_0 (t) + H_{\mm{int}}$, as the
one studied by LZS to derive the staying and transition probabilities $P_{\mm{a}}$ and $P_{\mm{na}}$
between two levels $\ket{1}$ and $\ket{2}$ driven through resonance between $t_{\mm{i}} =
-\infty$ and $t_{\mm{f}} =+\infty$ by assuming that their energy difference is a linear
function of time, $\Delta (t) = \left|E_{\mathrm{S},\mathbf{\chi}}(t) -
E_{\mathrm{T},\mathbf{\chi}}(t)\right| =\alpha t$ with the well-known result $P_{\mm{a}}
= 1- P_{\mm{na}} = 1 - \exp (-2 \pi |\langle 1|H_{\mm{HF}}|2\rangle|^2 /\alpha \hbar)$. 
Here, we study the effect that the transitions can have on the nuclear difference field
$\delta h^z = h_1^z - h_2^z$ and more precisely on its fluctuations $\sigma^{(z)}=
\sqrt{\left\langle (\delta h^z)^2\right\rangle - \left\langle \delta h^z
\right\rangle^2}$ that are responsible for the qubit decoherence via $T_2^* =
\hbar/\sigma^{(z)}$ \cite{hanson_review, footnote1}. 

To study the evolution of $\sigma^{(z)}$, we compute the new state of the system after
each cycle, which consists of a forward and a return sweep. The state of the system after
the forward sweep results from the time evolution operator generated by
\eqref{Lzhamiltonian}; the resulting state is taken as the initial condition for the
return sweep which is performed sufficiently fast to ensure a \textit{sudden} parameter
change without change in the state.  
As in the original LZS model, we assume a linear dependence $\Delta(t)=\alpha t$
throughout. Moreover, to make the problem treatable, we assume a constant hyperfine
coupling $A_k \rightarrow \bar{A}$ and the nuclear spins to be $1/2$ (in reality
the nuclear species of GaAs have spin $3/2$).
Before treating the case of many nuclear spins, we explain the main ideas by considering
the simple case of one nuclear spin in dot $1$ and none in dot $2$.
The initial nuclear state is assumed to be an incoherent mixture of spin up and down; its
density matrix $\rho^{(0)}_{\mm{n}}$ has matrix elements $\rho^{(0)}_{ij} = \delta_{ij}/2$ with
$i,j = \uparrow,\downarrow$. The initial mean value and standard deviation for $\delta
h^z$ are $\langle \delta h^z \rangle_0 = 0$ and $\sigma_0^{(z)} = \bar{A}/2$. We now
assume that after a cycle, a measurement of the electron spin (via charge) is performed
to determine if a flip-flop has occurred. If a singlet is detected, the nuclear density
matrix becomes $\rho^{(1),\mm{S}}_{\uparrow \uparrow} = e^{-\eta}/2P_\mm{S}$,
$\rho^{(1),\mm{S}}_{\downarrow \downarrow} = 1/2P_\mm{S}$, and $\rho^{(1),\mm{S}}_{\uparrow
\downarrow}=\rho^{(1),\mm{S}}_{\downarrow \uparrow}=0$, with $\eta = \pi \bar{A}^2
/\alpha \hbar$ and the probability to measure a singlet, $P_{\mm{S}} = (e^{-\eta} +
1)/2$. 
In this new state, $\langle \delta h^z \rangle_1 = \bar{A} (e^{-\eta} -1) / 2 (e^{-\eta}
+ 1)$ and $\sigma_1^{(z)} = (\bar{A}/2)\cosh^{-1}(\eta/2)$. In the ``fast'' limit $\alpha \to
\infty$, we find $\rho^{(1),\mm{S}}_{\mm{n}} = \rho^{(0)}_{\mm{n}}$ therefore the
variance is unchanged, $\sigma_1^{(z)} = \sigma_0^{(z)}$. In the ``slow'' limit $\alpha
\to 0$, we have $\rho^{(1),\mm{S}}_{\downarrow \downarrow} = 1$ while all other elements
of $\rho^{(1),\mm{S}}_{\mm{n}}$ vanish, such that $\sigma_1^{(z)} = 0$, describing a
strong (projective) measurement of the nuclear spin.  For $0<\alpha <\infty$, the
detection of a singlet induces a \emph{weak} measurement which decreases the
fluctuations, $\sigma_1^{(z)} < \sigma_0^{(z)}$. 
On the other hand, if a triplet is measured, we have $\rho^{(1),\mm{T}}_{\uparrow
\uparrow} = \rho^{(1),\mm{T}}_{\uparrow \downarrow} = \rho^{(1),\mm{T}}_{\downarrow
\uparrow} =0$ and $\rho^{(1),\mm{T}}_{\downarrow \downarrow}=1$ independently of
$\alpha$; in this case the nuclear spin is projected on the down state. 
We conclude that it is possible to reduce the fluctuations of the nuclear spins without
fully polarizing them.  
Below, we show that the same mechanism works also for a system with many nuclear
spins. 

For $n \gg 1$, the nuclear states are highly degenerate and the LZS propagator from the
simple case cannot be used anymore. An elegant solution to derive the LZS propagator for
degenerate systems  consists in applying the unitary Morris-Shore (MS) transformation to
the LZS Hamiltonian to reduce the dynamics into sets of decoupled single states and
independent two-level systems \cite{vasilev_deg_lz}. Because $H_{\mm{HF}}$ exclusively
couples states of the form $\ket{\mathrm{S},j_1, m_1, j_2, m_2} \equiv \ket{0}$ to the
degenerate states $\ket{\mathrm{T_{+}},j_1, m_1 -1, j_2, m_2}\equiv \ket{1}$ and
$\ket{\mathrm{T_{+}},j_1, m_1, j_2, m_2 -1} \equiv \ket{2}$, the LZS Hamiltonian can be
brought into a block diagonal form. The MS basis is found by diagonalizing
\begin{equation} \mathbf{V}\mathbf{V}^\dag = \frac{\bar{A}^2}{8} \begin{pmatrix} a_1^2 +
a_2^2 & 0 & 0\\ 0 & a_1^2 & a_1 a_2\\ 0 & a_1 a_2 & a_2^2 \\ \end{pmatrix},
\label{ms_int_matrix} \end{equation} where $V_{ij}=\langle i| H_{\rm int}|j\rangle$, $a_i
= \sqrt{j_i (j_i + 1) - m_i (m_i -1)}$. The eigenstates are $\ket{0}$ and
$\ket{1'}=\left(a_1 \ket{1} + a_2 \ket{2}\right)/a_0$ associated with the eigenvalue
$\lambda^2=\bar{A}^2 /8 a_0^2$, and $\ket{2'}=\left(a_1 \ket{1} - a_2 \ket{2}\right)/a_0$
with eigenvalue $0$, where $a_0^2 = a_1^2 + a_2^2$. The states with the
same eigenvalues are coupled with strength $\lambda$. The state $\ket{2'}$
is a ``dark state'', as it does not couple to other states.
In the subspace spanned by $\ket{0}$ and $\ket{1'}$, the time-dependent Schr\"odinger
equation with the initial state $\ket{0}$ at time $t_{\mm{i}}$ can be solved and thus the
LZS propagator elements $U_{0\,i'}^{\mm{MS}}$ can be calculated \cite{vitanov_time_lz}.
In order to express the solution in the original basis, we perform the inverse MS
transformation to find the matrix elements $U_{0\,i} =(a_i/a_0)U_{0\,i'}^{\mm{MS}}$.
We also account for finite time propagation and thus avoid the unphysical situations of
infinite energy that arises for couplings that do not vanish when $t \rightarrow \pm
\infty$, and infinite detuning as we assume $\Delta (t) = \alpha t$. The finite-time
solution also allows us to model the situation in the experiment \cite{reilly_science}
where mixing between $\mathrm{T}_0 (1,\,1)$ and the $\mathrm{S} (1,\,1)$ states must be
avoided.

At typical operating temperatures and fields, where $k_B T\gg g_{\mm{N}}\mu_{\mm{N}} B$,
the initial nuclear density matrix can be assumed to be diagonal,
$\rho_{\mathrm{n}}^{(0)} =
\sum_{\mathbf{\chi}}p(\mathbf{\chi})\ket{\mathbf{\chi}}\bra{\mathbf{\chi}}$ with a
uniform distribution of states $\chi$.
At $120\,\mathrm{mK}$ and $100\,{\rm mT}$, we have $k_{\mathrm{B}}T \sim
10^{-5}\,\mathrm{eV}$ and  $g_{\mathrm{N}}\mu_{\mathrm{N}}B\sim 10^{-9}\,\mathrm{eV}$.
The joint probability $p(\mathbf{\chi})$ can be factorized into $p'(j_1, m_1) p'(j_2,
m_2)$ since the dots are initially independent, with $p'(j_i, m_i) = g(j_i) f(m_i|j_i)$, where
$g(j_i)$ is the probability for total nuclear angular momentum $j_i$ and $f(m_i|j_i) =
[\theta(j_i + m_i) -\theta(j_i - m_i)]/(2 j_i + 1)$ is the equally distributed
conditional probability of having a magnetization $m_i$ given $j_i$, and $\theta$ is the
Heaviside function with $\theta (0)=1$. The probability distribution $g(j_i) =
G(j_i)/\sum_{j_i}G(j_i)$ is found by counting how many times $G(j_i)$ an irreducible
representation of dimension $2 j_i + 1$ occurs.

After the forward sweep of the $k$-th cycle, the state is
$\rho^{(k+1)} = U \rho^{(k)} U^{\dag}$, with $\rho^{(0)} =
\ket{\mm{S}}\bra{\mm{S}}\otimes \rho^{(0)}_{\mm{n}}$. 
The back sweep will act as a measurement of the final configuration of the electronic
system. After the LZS transition the charge configuration of the system is $(1,1)$
independent of the spin, but after the back sweep it is a superposition of
$\mm{T}_+ (1,1)$ and $\mm{S}(2,0)$ which has a relatively fast decay time $\tau\sim
1\,\mm{ns}$ \cite{hayashi_fujisawa}, such that the electronic system will evolve with
probability $P_{\mm{S}}$ to the singlet $\mm{S}(2,0)$ and with probability $P_{\mm{T}}$
to the triplet $\mm{T}_+ (1,1)$ after a time $\tau$.  
This provides a way to determine if the system has evolved adiabatically or not during
the forward sweep. 
We write $\rho^{(k+1)} = P_{\mm{S}} \rho_{\mm{S}}^{(k+1)} +
P_{\mm{T}}\rho_{\mm{T}}^{(k+1)}$ to describe the mixture of the ensembles that have
evolved adiabatically and non-adiabatically in the forward sweep. After a time $\tau$,
this state will collapse either to $\rho_{\mm{S}}^{(k+1)}$ or $\rho_{\mm{T}}^{(k+1)}$
with probability $P_{\mm{S}}$ and $P_{\mm{T}}$, respectively, with
($q=\mm{S},\mm{T}$),
\begin{equation} 
P_q = \mathrm{Tr}\left[ M_q U \rho^{(k)}
U^{\dag} M_q^{\dag} \right] ,
\label{stay_prob} 
\end{equation}
where $M_q=\ket{q}\bra{q}$ is the projection operator describing a strong
measurement in the charge sector.
If a singlet ($q=\mm{S}$) or triplet ($q=\mm{T}$) is detected after the $(k+1)$-th sweep,
we update $\rho_{\mm{n}}$ according to
\begin{equation} 
\rho_{\mathrm{n}}^{(k+1)} =
\frac{1}{P_q}
M_{q} U \rho_{\mathrm{n}}^{(k)} U^{\dag}M_{q}^{\dag}.
\label{update_rule} 
\end{equation}
In the case of a triplet ($q=\mm{T}$), we must also take into account the nuclear spin flop,
\begin{equation}
\begin{aligned}
\rho^{(k+1)}_{j_1  j_2  j_1  j_2} \rightarrow &0,\\
\rho_{j_1  j_2  m_1  m_2}^{(k+1)} \rightarrow & \frac{\rho_{j_1  j_2,  m_1 + 1, m_2}^{(k+1)} +
\rho_{j_1  j_2  m_1, m_2 + 1}^{(k+1)}}{\sum_{j_1 j_2 m_1 m_2} \left(\rho_{j_1  j_2,
m_1 + 1, m_2}^{(k+1)} + \rho_{j_1 j_2 m_1, m_2+1}^{(k+1)}\right)}.
\end{aligned}
\label{update_rule_triplet_2nd_step}
\end{equation}
After a flip-flop process, the crossing point $\varepsilon^{*}$ moves slightly because of
the change in the total magnetic field $\mathbf{B} + (\mathbf{h}_1 + \mathbf{h}_2)/g^{*}
\mu_{\mm{B}}$ and affects the initial and final times of the LZS transitions. With the
convention that $t=0$ at the initial anti-crossing point, the time shift is $
t_{\mathrm{i}}\rightarrow t_{\mathrm{i}} + \delta t$, $t_{\mathrm{f}}\rightarrow
t_{\mathrm{f}} + \delta t$, with $\delta t = \bar{A} / \alpha$.

We have performed a Monte Carlo simulation where $t_{\mathrm{i}} = -30\,\mathrm{ns}$ and
$t_{\mathrm{f}} = 20\,\mathrm{ns}$, such that the duration of the first sweep is the same
as in \cite{reilly_science}. Our choice for $\alpha$ satisfies the conditions
$\lambda \ll \alpha t_{\mathrm{i}}$ and $\lambda \ll \alpha t_{\mathrm{f}} \ll g^{*}
\mu_{\mathrm{B}} B$ that hold for a system far from resonance at the beginning and at the
end of the first cycles, and with a final difference in energy smaller than the Zeeman
splitting. The order of magnitude of $\lambda$ is determined by looking at the case of
the most likely $j_i$, which is $\sim \sqrt{n}$, with $m_i = 0$ to consider the strongest
coupling in that subspace. The number of spins is currently limited to $n=200$ \emph{per
dot} by the computational power at our disposal. This limitation has a side effect on the
choice of the hyperfine coupling constant. For large spin systems, one has $\bar{A} =
\frac{A}{n}$ with $A = 90\,\mathrm{\mu eV}$ for GaAs, but for dilute systems $\bar{A}$
can be crucially smaller, therefore  we choose $\bar{A} = 9\,\mathrm{neV}$
\cite{footnote2}. Thus, $\lambda \sim 10^{-8}\,\mm{eV}$ so that $\alpha t_{\mathrm{f}}$
must be of the order of $10^{-7}\,\mathrm{eV}$ since $\left|g^{*} \mu_{\mm{B}} B\right|
\simeq 2.3\,\mm{\mu eV}$. This implies that $\alpha$ must be between $10$ and
$50\,\mathrm{eV\,s^{-1}}$, which also satisfies the condition on $\alpha t_{\mathrm{i}}$.

\begin{figure} 
\includegraphics[width=.48\textwidth]{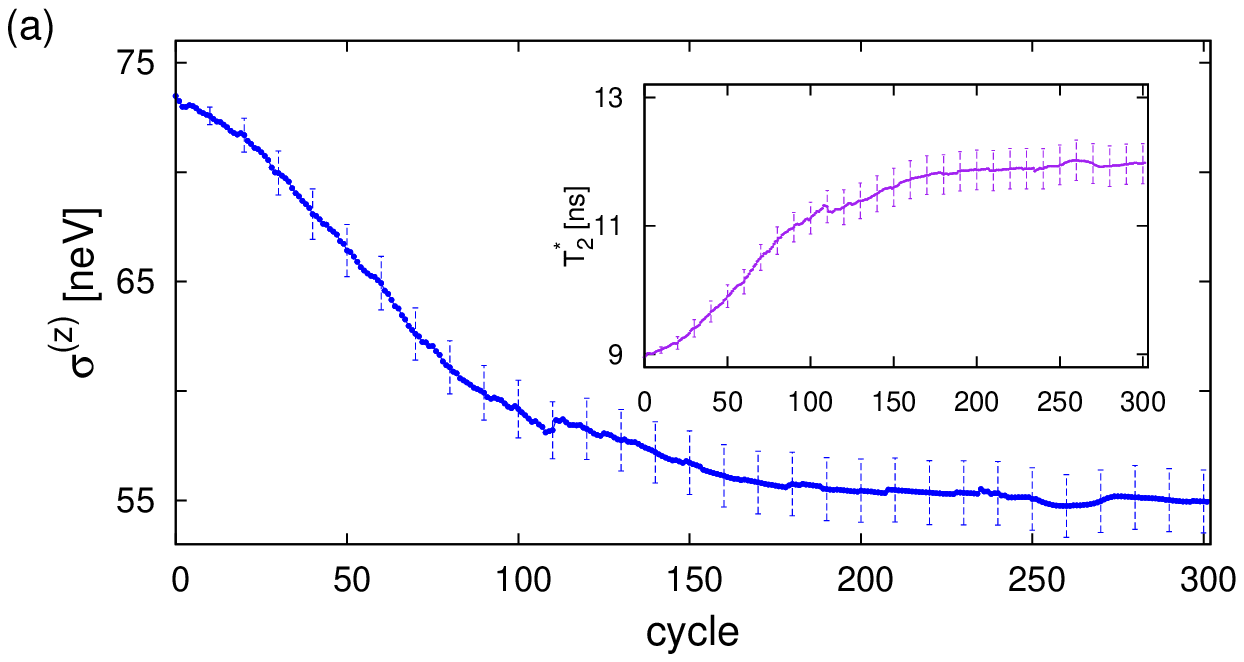}
\includegraphics[width=.48\textwidth]{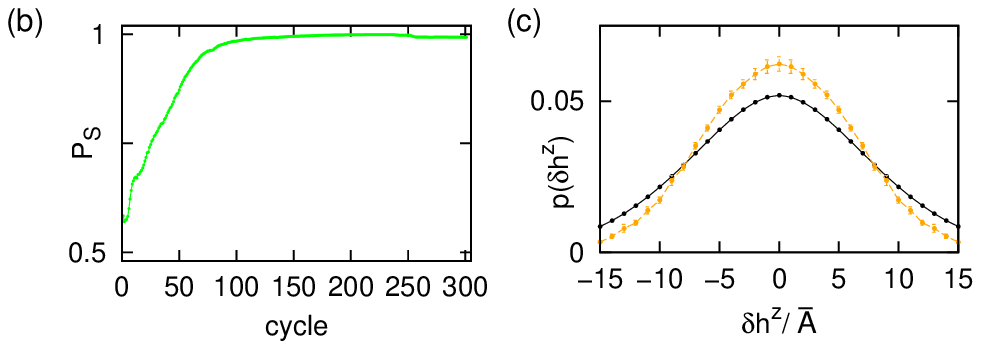} 
\caption{ (color online)
(a) Averaged Overhauser field fluctuations $\sigma^{(z)}$ and electron spin decoherence 
$T_2^* = \hbar/\sigma^{(z)}$ (upper inset) as a function of the number of performed cycles.
(b) Singlet return probability $P_{\mm{S}}$ as a function of the number of performed
cycles. Error bars are smaller than line thickness.
(c) Initial (black) and final (orange) averaged probability distribution of the nuclear
spin eigenstates of $\delta h^z$. Repeated cycles narrow the distribution, showing a
reduction of $\sigma^{(z)}$.}
\label{plot_results} 
\end{figure}

\textit{Results---}
In Fig.~\ref{plot_results}(a), we plot the evolution of $\sigma^{(z)}$, averaged over
$160$ runs, as a function of the number of cycles for $\alpha = 11 \,\mathrm{eV\,s^{-1}}$.
Since the reduction of $\sigma^{(z)}$ persists in the average, a read-out of the charge state
after each cycle is not required.
When the polarization reaches about $8\%$ (generally at $\approx 1/\sqrt{n/2}$) the
nuclear spins approach a ``dark state'' preventing further polarization. We plot in
Fig.~\ref{plot_results}(b) the average probability $P_{\mm{S}}$ of measuring a singlet as
a function of the number of cycles. $P_{\mm{S}}$ can be measured in the $\mm{S}$ -
$\mm{T}_+$, as it was done for $\mm{S}$ - $\mm{T}_0 $ \cite{petta_science}.

\textit{Conclusion---}
We have developed a model to explain the increase of $T^{*}_2$ in double
QDs via the tuning of two electrons across an energy level anti-crossing between
$\mm{S}(2,0)$ and $\mm{T}_+ (1,1)$. Our model is based on the possibility of
\emph{partially adiabatic} transitions which are described by a generalized LZS theory. We
have shown that the cycling combined with a spin-to-charge conversion to the $(1,1)$,
respectively $(2,0)$, configuration induces a \emph{weak measurement} on the nuclear state
which strongly contributes in the suppression of the nuclear fluctuations. An
experimental confirmation of the predicted \emph{weak measurement} would be provided by
measuring $P_{\mm{S}} \approx 1$ and at the same time an increase in $T_2^*$ as a
function of the number of cycles.

Using a Monte-Carlo algorithm for $n=200$ nuclear spins per QD, we find an enhancement of
the electron spin coherence time $T_2^* = \hbar / \sigma^{(z)}$ by a factor of $\sim
1.34$ after 300 cycles.
While the initial $\sigma_{\mm{i}}^{(z)} \propto \sqrt{n}$, the final
$\sigma_{\mm{f}}^{(z)}$ is mainly determined by the preparation mechanism and is
approximatively independent of $n$. Therefore, we expect $T_{2,\mm{f}}^* / T_{2,\mm{i}}^*
= \sigma_{\mm{i}}^{(z)} / \sigma_{\mm{f}}^{(z)} \propto \sqrt{n}$, and we estimate
$T_{2,\mm{f}}^* / T_{2,\mm{i}}^* \approx 94$ for $n \approx 10^6$. This scaling stops
when $T_2^* \approx T_2$ or before if dipolar interaction effects are taken into account.
Future calculations for more spins will allow more direct comparison with experiments.
Our results also apply for QDs with fewer nuclear spins e.g. in Si, ZnO,
carbon nanotubes, or graphene.

%\section*{Acknowledgments}
\textit{Acknowledgments---}
We thank M. Braun and A. Romito for useful discussions. We acknowledge funding from the DFG within SPP
1285 ``Spintronics'' and FOR 912, and from the Swiss SNF via grant no. PP002-106310.


\begin{thebibliography}{99}

\bibitem{loss_divicenzo_proposaL_QD}
D. Loss, D. P. DiVincenzo, Phys. Rev. A \textbf{57}, 120 (1998).

\bibitem{hanson_review}
R. Hanson \emph{et al.}, Rev. Mod. Phys \textbf{79}, 1217 (2007).

\bibitem{khaetskii_e-spin_decoherence}
A. V. Khaetskii, D. Loss, L. Glazman, Phys. Rev. Lett. \textbf{88}, 186802 (2002).

\bibitem{levy_QC_2e}
J. Levy, Phys. Rev. Lett. \textbf{89}, 147902 (2002).

\bibitem{taylor_nature_phys}
J. M. Taylor \emph{et al.}, Nature Physics \textbf{1}, 177 (2005).

\bibitem{hanson_burkard}
R. Hanson, G. Burkard, 
Phys.\ Rev.\ Lett. \textbf{98}, 050502 (2007).

\bibitem{coish_s-t_decoherence}
W. A. Coish, D. Loss, Phys. Rev. B \textbf{72}, 125337 (2005).

\bibitem{petta_science}
J. R. Petta \emph{et al.}, Science \textbf{309}, 2180 (2005).

\bibitem{klauser_nucl_spin_narr}
D. Klauser, W. A. Coish, D. Loss, Phys. Rev. B \textbf{73}, 205302 (2006).

\bibitem{stepanenko_enh_spin_coh_opt}
D. Stepanenko, G. Burkard, G. Giedke, A. Imamoglu, Phys. Rev. Lett. \textbf{96} 136401,
(2006).

\bibitem{burkard_polarization}
G. Burkard, D. Loss, D. P. DiVincenzo, Phys. Rev. B \textbf{59}, 2070 (1999).

\bibitem{petta_polarization}
J. R. Petta \emph{et al.}, Phys. Rev. Lett. \textbf{100}, 067601 (2008).

\bibitem{coish_99}
W. A. Coish, D. Loss, Phys. Rev. B \textbf{70}, 195340 (2004).

\bibitem{reilly_science}
D. J. Reilly \emph{et al.}, Science \textbf{321}, 817 (2008).

\bibitem{ramon_hu}
G. Ramon, X. Hu, Phys. Rev. B \textbf{75}, 161301 (2007).

\bibitem{aharonov_wm}
Y. Aharonov, L. Vaidman, quant-ph/0105101v2 (2007).

\bibitem{landau32}
L. D. Landau, Phys.\ Z.\ Sowjetunion {\bf 2}, 46 (1932). 

\bibitem{zener32}
C. Zener, Proc.\ R.\ Soc.\ A {\bf 137}, 696 (1932). 

\bibitem{stuckelberg32}
E. C. G. St\"uckelberg, Helv.\ Phys.\ Acta {\bf 5}, 369 (1932). 

\bibitem{vasilev_deg_lz}
G. S. Vasilev, S. S. Ivanov, N. V. Vitanov, Phys. Rev. A \textbf{75}, 013417 (2007).

\bibitem{vitanov_time_lz}
N. V. Vitanov, B. M. Garraway, Phys. Rev. A \textbf{53}, 4288 (1996).

\bibitem{hayashi_fujisawa}
T. Hayashi \emph{et al.}, Phys. Rev. Lett. \textbf{91}, 226804 (2003).

\bibitem{footnote1}
Writing $H_{\mathrm{HF}} =
(\mathbf{S}\cdot\mathbf{h}+\delta\mathbf{S}\cdot\delta\mathbf{h})/2$, with $\mathbf{x} =
\mathbf{x}_1 + \mathbf{x}_2$ and $\delta\mathbf{x} = \mathbf{x}_1 - \mathbf{x}_2$
($\mathbf{x}_i = \mathbf{S}_i,\,\mathbf{h}_i,\,i=1,2$), one finds that only the difference operators
lead to the decoherence ($T_2^{*}$) of the $\mm{S}-\mm{T}_0$ qubit.

\bibitem{footnote2}
The choice of $\bar{A}$ is made such that the most likely $\lambda$ is the same for $n=200$ and $n=10^6$.

\end{thebibliography}
\end{document}